\documentclass[journal=nalefd,manuscript=article]{achemso}

\usepackage[version=3]{mhchem} 

\SectionNumbersOn    

\usepackage[capitalize]{cleveref}
\crefname{figure}{Fig.}{Figs.}
\Crefname{figure}{Figure}{Figures}
\crefname{table}{Tab.}{Tabs.}
\Crefname{table}{Table}{Tables}
\crefname{equation}{Eq.}{Eqs.}
\Crefname{equation}{Equation}{Equations}
\crefname{section}{Sec.}{Secs.}
\Crefname{section}{Section}{Sections}

\usepackage{subcaption}
\usepackage{siunitx, soul, xcolor}
\newsavebox{\measurebox}
\usepackage{gensymb}

\usepackage[export]{adjustbox}

\DeclareCaptionFont{capt}{\fontseries{n}\fontfamily{phv}\selectfont}

\let\oldst\st
\renewcommand{\st}[1]{{\textcolor{blue}{\oldst{#1}}}}

\usepackage{subfiles} 

\author{Wenyi Zhou}
\affiliation{The Ohio State University, Department of Physics, Columbus, Ohio 43210, USA}
\author{Alexander J. Bishop}
\affiliation{The Ohio State University, Department of Physics, Columbus, Ohio 43210, USA}
\author{Menglin Zhu}
\affiliation{The Ohio State University, Department of Materials Science and Engineering, Columbus, Ohio 43210, USA}
\author{Igor Lyalin}
\affiliation{The Ohio State University, Department of Physics, Columbus, Ohio 43210, USA}
\author{Robert C. Walko}
\affiliation{The Ohio State University, Department of Physics, Columbus, Ohio 43210, USA}
\author{Jay A. Gupta}
\affiliation{The Ohio State University, Department of Physics, Columbus, Ohio 43210, USA}
\author{Jinwoo Hwang}
\affiliation{The Ohio State University, Department of Materials Science and Engineering, Columbus, Ohio 43210, USA}
\affiliation{The Ohio State University, Department of Physics, Columbus, Ohio 43210, USA}
\author{Roland K. Kawakami}
\affiliation{The Ohio State University, Department of Physics, Columbus, Ohio 43210, USA}
\email{kawakami.15@osu.edu}

\title{Kinetically-controlled epitaxial growth of Fe$_3$GeTe$_2$ van der Waals ferromagnetic films}


\begin{document}


\begin{abstract}
We demonstrate that kinetics play an important role in the epitaxial growth of Fe$_3$GeTe$_2$ (FGT) van der Waals (vdW) ferromagnetic films by molecular beam epitaxy. 
By varying the deposition rate, we control the formation or suppression of an initial tellurium-deficient non-van der Waals phase (Fe$_3$Ge$_2$) prior to realizing epitaxial growth of the vdW FGT phase.
Using cross-sectional scanning transmission electron microscopy and scanning tunneling microscopy, we optimize the FGT films to have atomically smooth surfaces and abrupt interfaces with the Ge(111) substrate. The magnetic properties of our high quality material are confirmed through magneto-optic, magnetotransport, and spin-polarized STM studies. Importantly, this demonstrates how the interplay of energetics and kinetics can help tune the re-evaporation rate of chalcogen atoms and interdiffusion from the underlayer, which paves the way for future studies of van der Waals epitaxy.
\end{abstract}

\section*{Introduction}

\begin{figure}
  \adjustbox{minipage=1.1em,valign=t}{\subcaption{}\label{fig:FGT_atomic_structure}}
  \begin{subfigure}[t]{\dimexpr.45\textwidth-1.1em\relax}
  \centering
    \includegraphics[width=\textwidth,valign=t]{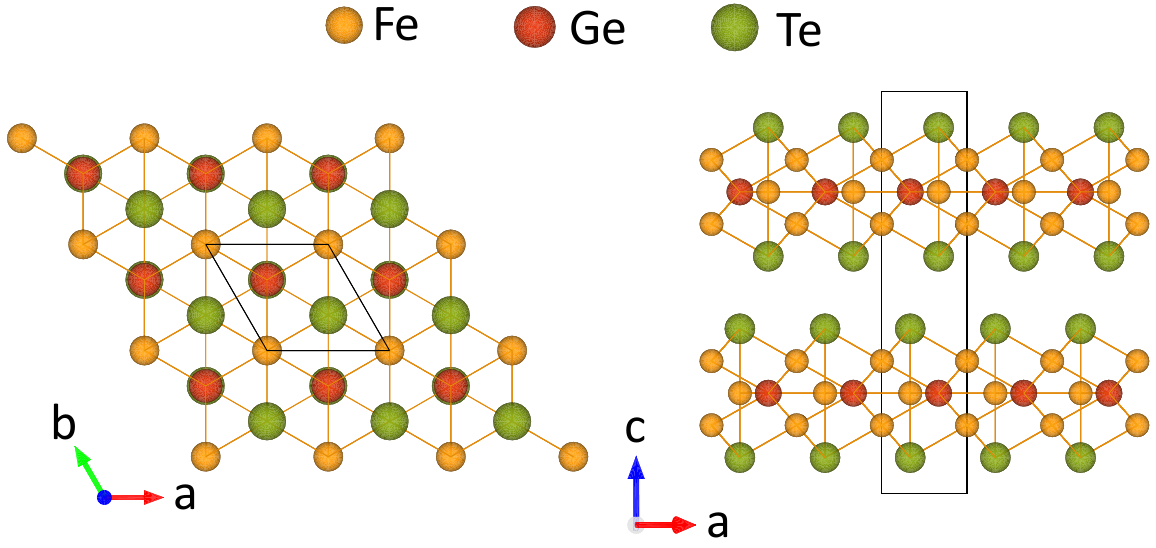}
    \end{subfigure}%
\hfill
  \adjustbox{minipage=0.9em,valign=t}{\subcaption{}\label{fig:Ge_RHEED}}%
  \begin{subfigure}[t]{\dimexpr.45\textwidth-0.9em\relax}
  \centering
    \includegraphics[width=0.95\textwidth,valign=t]{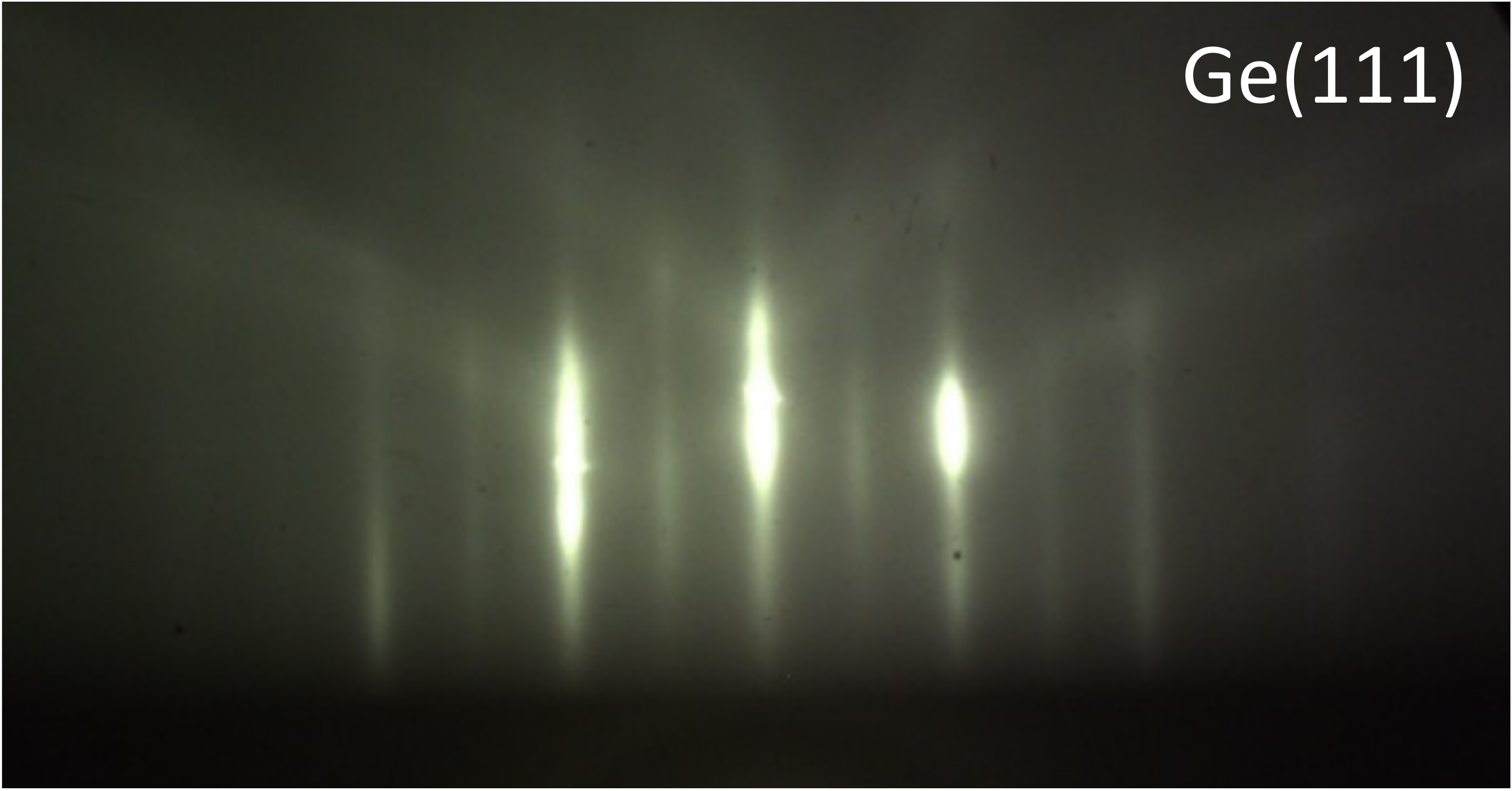}
    \end{subfigure}
\hfill
\adjustbox{minipage=0.9em,valign=t}{\subcaption{}\label{fig:slow_FGT_RHEED}}%
  \begin{subfigure}[t]{\dimexpr.45\textwidth-0.9em\relax}
  \centering
    \includegraphics[width=0.95\textwidth,valign=t]{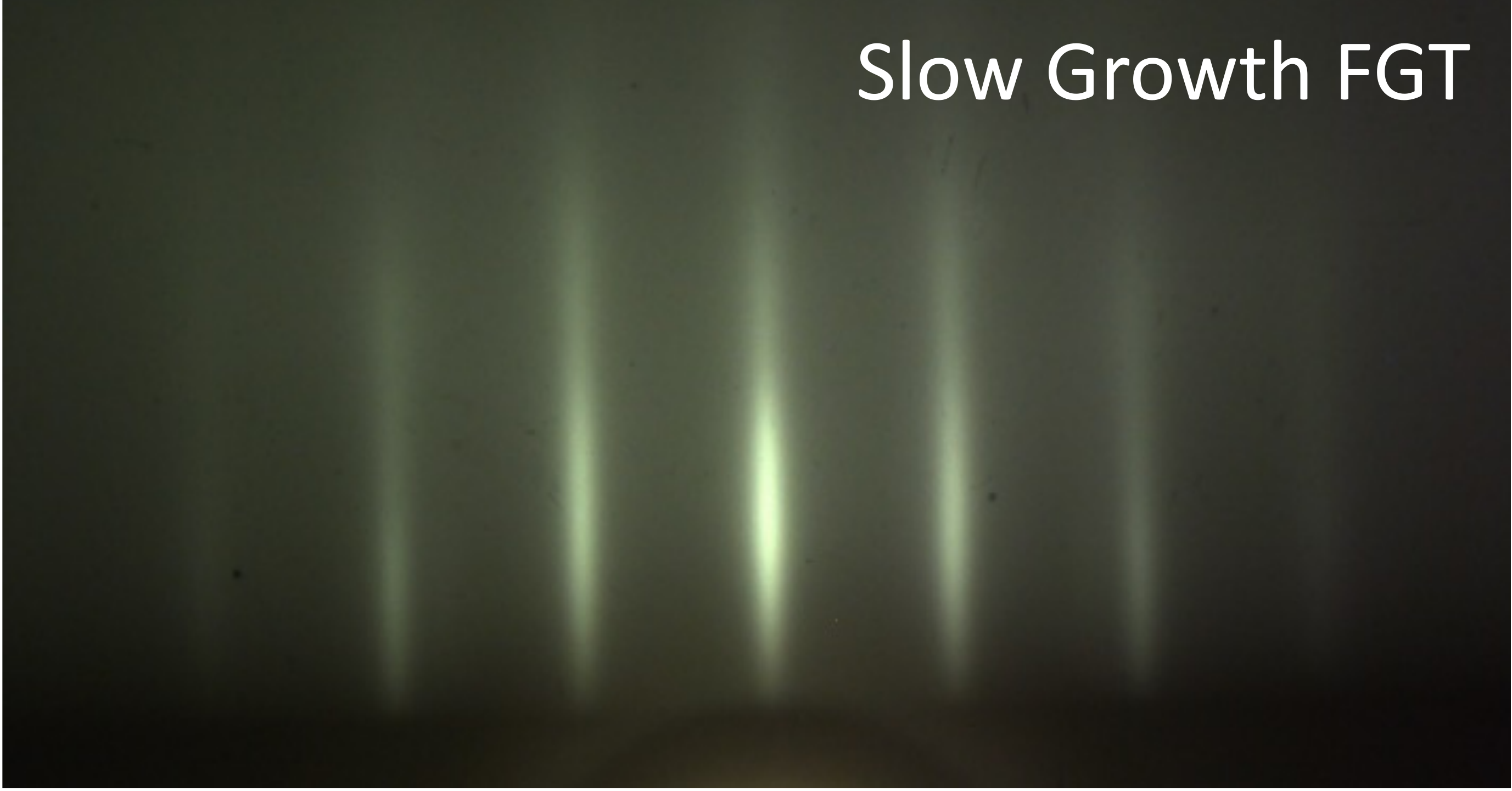}
  \end{subfigure}
  \hfill
\adjustbox{minipage=0.9em,valign=t}{\subcaption{}\label{fig:fast_FGT_RHEED}}%
  \begin{subfigure}[t]{\dimexpr.45\textwidth-0.9em\relax}
    \centering
    \includegraphics[width=0.95\textwidth,valign=t]{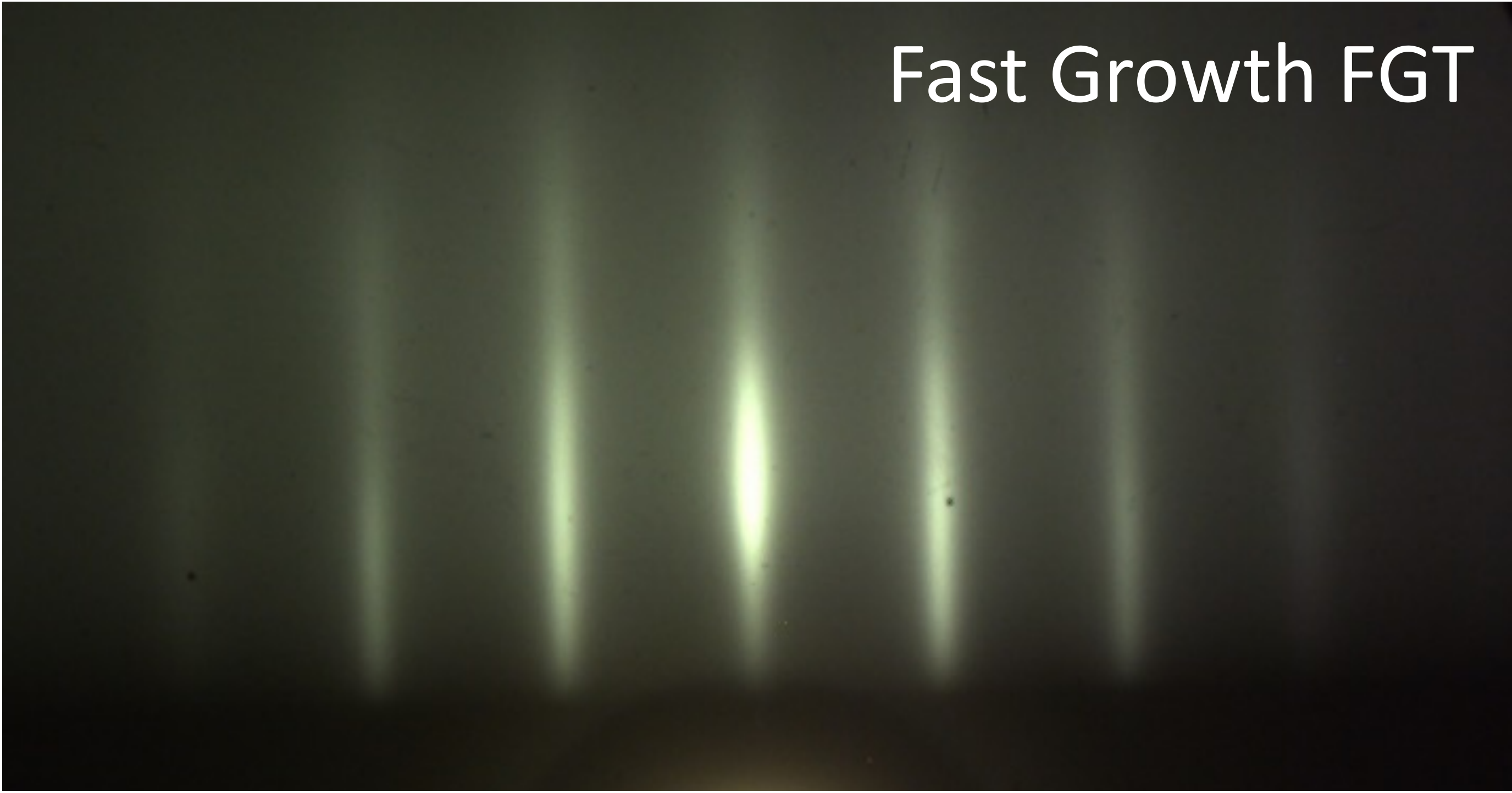}
  \end{subfigure}
  \hfill
\adjustbox{minipage=0.9em,valign=t}{\subcaption{}\label{fig:slow_FGT_XRD}}%
  \begin{subfigure}[t]{\dimexpr.45\textwidth-0.9em\relax}
    \centering
    \includegraphics[width=\textwidth,valign=t]{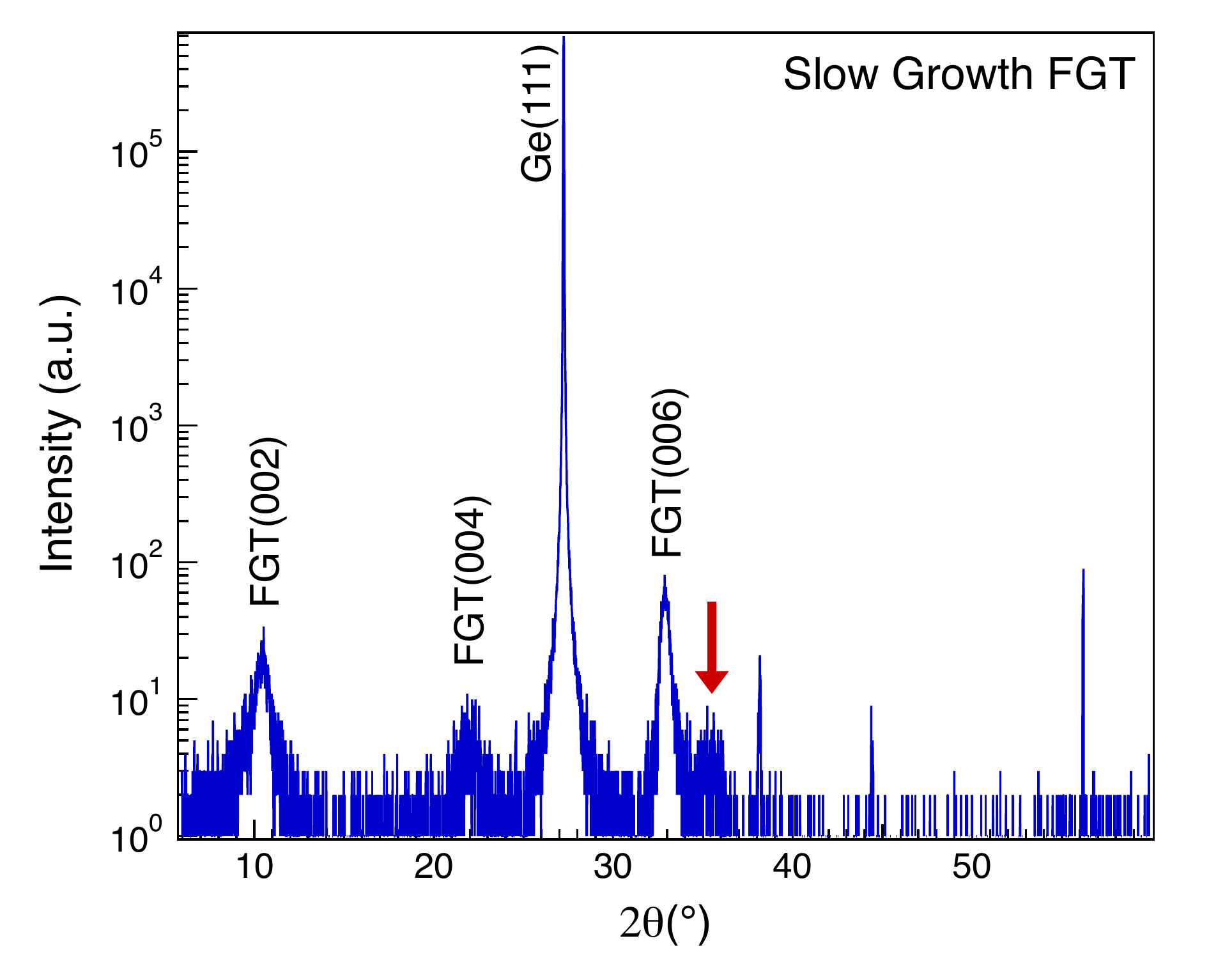}
  \end{subfigure}
  \hfill
\adjustbox{minipage=0.9em,valign=t}{\subcaption{}\label{fig:fast_FGT_XRD}}%
  \begin{subfigure}[t]{\dimexpr.45\textwidth-0.9em\relax}
    \centering
    \includegraphics[width=\textwidth,valign=t]{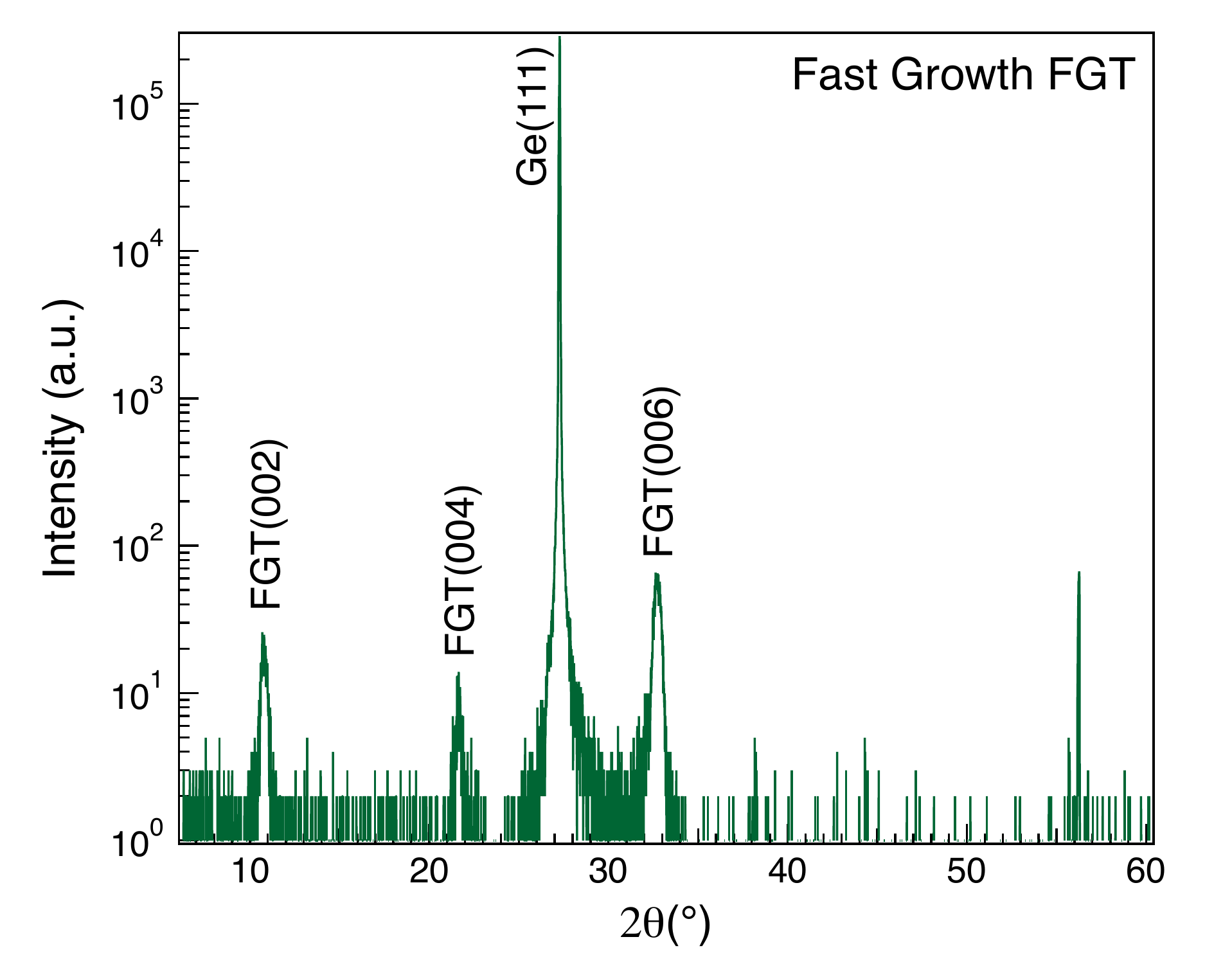}
  \end{subfigure}
  \hfill
   \caption{MBE-grown Fe$_3$GeTe$_2$ (FGT) films on Ge(111) at different growth rates. (a) Atomic
lattice of FGT, where the crystal unit cell is outlined. (b) RHEED pattern of Ge(111) substrate. (c) RHEED
pattern for 40 nm FGT grown at 0.06 \AA /s (\textquotedblleft Slow Growth
FGT\textquotedblright ). (d) RHEED pattern for 20 nm FGT grown at
0.12 \AA /s (\textquotedblleft Fast Growth FGT\textquotedblright ). (e) $\theta-2\theta$
x-ray diffraction scan of Slow Growth FGT (40 nm). The red arrow indicates the presence
of an unexpected additional peak. (f) $\theta-2\theta$
x-ray diffraction scan of Fast Growth FGT (20 nm), which does not exhibit the additional
peak. \label{fig:RHEED_and_XRD}
}
\end{figure}

There is tremendous interest in ultrathin van der Waals magnets due to the unprecedented tunability of magnetic properties via strain\cite{xiaodong2022,zhuoliang2021,YuWang2020}, electrostatic gating\cite{jiang2018,huang2018}, and incorporation into heterostructures\cite{song2018,klein2018,wang2018,zhong2020}. Fe$_3$GeTe$_2$ (FGT) (Figure~\ref{fig:FGT_atomic_structure}) is one of the most well-studied vdW magnets due to its strong perpendicular magnetic anisotropy (PMA), relatively high Curie temperature ($T_C$) ranging from $\sim$200 K to above room temperature\cite{fei2018,deng2018,li2018,wang2020RT,alghamdi2019,kao2020}, and observation of novel phenomena such as spin-orbit torque switching\cite{wang2019,alghamdi2019,kao2020,wang_room_2021} and skyrmion spin textures\cite{wu_ne-type_2020,yang_creation_2020,Tae2021}.

While many efforts have focused on small flakes exfoliated from bulk crystals, epitaxial growth of thin films and heterostructures by methods such as molecular beam epitaxy (MBE) provide unique opportunities. This includes stabilizing materials far from equilibrium that cannot be realized in bulk crystal form (e.g. vdW MnBi$_2$Se$_4$\cite{zhu2021}), performing atomic-scale manipulation of interface structure and material properties, and enabling scaling up to large areas for electronic and magnetic applications.

To date, there have been several reports of epitaxial growth of FGT and related materials by MBE (e.g. FGT films on Al$_2$O$_3$(0001)\cite{liu2017}, GaAs(111)\cite{liu2017}, Ge(111)\cite{roemer_robust_2020}, and Bi$_2$Te$_3$(0001)\cite{wang2020RT,chen_generation_2021,wang_room_2021}; Fe$_5$GeTe$_2$ on Al$_2$O$_3$(0001)\cite{MarioRibeiro_F5GT_2021}). However, critical issues of the growth process such as the relative importance of thermodynamic equilibrium (energetics) and growth/reaction rates (kinetics) remain largely unexplored. Understanding these issues are important for improving the material quality, as it relates to important properties of thin films and heterostructures including interdiffusion and defect formation.

In this work, we demonstrate that kinetics play an important role in the epitaxial growth of FGT films by MBE. 
For FGT grown on Ge(111), we find that growth rate is a critical parameter that determines whether FGT grows directly on Ge(111) with an atomically abrupt interface or whether a Te-deficient non-vdW Fe$_3$Ge$_2$ alloy forms at the interface. Contrary to expectations for reactive MBE growth\cite{schlom_oxide_2001,theis_adsorption-controlled_1998}, our studies show that the Fe$_3$Ge$_2$ alloy forms at lower growth rates while the abrupt FGT/Ge(111) interface is obtained for higher growth rates. This behavior suggests that energetic considerations favor the formation of non-vdW Fe-Ge alloys, while the growth kinetics can favor the stabilization of pure FGT. The high quality of the pure FGT films is corroborated by their excellent magnetic properties, where $\sim$20 nm FGT films on Ge(111) exhibit typical properties of bulk FGT including perpendicular magnetic anisotropy with square hysteresis loops and a $T_C$ of $\sim$250 K. These results demonstrate that the interplay of energetics and kinetics can help tune the re-evaporation rate of chalcogen atoms and interdiffusion from the underlayer to facilitate the development of high quality vdW epitaxial films.

\section*{Experimental Section}

The FGT films were deposited in a Veeco 930 MBE system with base pressure of $2 \times 10^{-10}$ Torr. The Ge(111) substrates (MTI Corporation) were prepared by sonication in acetone and isopropyl alcohol for 5 min each before being inserted into the MBE chamber. After degassing at 800 \degree C for 12 min, the Ge substrates were cooled down to 325 \degree C for growth of the FGT films.
For synthesis, the films were grown by co-depositing Fe (99.99\%, Alfa Aesar), Ge (99.9999\%, Alfa Aesar), and Te (99.9999\%, United Mineral Corp.) with an atomic flux ratio of 3:1:20, as measured by a beam flux monitor (BFM) and calibrated by x-ray reflectivity (XRR) of elemental films. Prior to growth, the substrate was exposed to the Te flux for 5 minutes.

We employed a wide range of characterizations including \textit{in situ} by reflection high energy electron diffraction (RHEED) and \textit{ex situ} by x-ray diffraction to probe the surface and bulk crystal structure, respectively. The atomic-scale structure was investigated by scanning tunneling microscopy (STM) and cross-sectional scanning transmission electron microscopy (STEM) with elemental analysis (i.e.~energy dispersive x-ray spectroscopy (EDX)). Magnetic properties were investigated by anomalous Hall effect (AHE), optical magnetic circular dichroism (MCD), and spin-polarized STM.

For cross-sectional STEM imaging, specimens were prepared using conventional focused ion beam lift-out technique with Helios NanoLab 600. Argon ion milling under 900 and 500 eV was used subsequently to clean the surface amorphous layer and minimize subsurface damage. High angle annular dark field (HAADF) STEM images and EDX maps were collected using a probe-aberration corrected Thermo Fisher Scientific Themis Z S/TEM operated at 300 kV, 20 mrad convergence semi-angle.

For STM measurements, samples mounted on flag-style sample holders were transferred from the MBE chamber to the STM using a UHV suitcase to keep the sample under vacuum during the entire sample transfer. The STM measurements were performed in a Createc LT-STM/AFM system with the sample held at a temperature of 5 K. Spin-polarized STM measurements utilized a Cr tip and an out-of-plane magnetic field.

We performed AHE and MCD measurements in an Oxford Spectromag magneto-optical cryostat. To obtain out-of-plane hystersis loops, we applied a magnetic field perpendicular to the surface of the sample. The AHE measurements were performed using lock-in detection at 987\,Hz and an excitation current of 100\,$\mu$A (rms).
For MCD measurements, we utilized 
a 100\,$\mu$W, 532\,nm laser beam focused to a spot size of $\sim$150\,$\mu$m. The helicity of the incident beam was modulated at 50 kHz by a photoelastic modulator and the MCD of the reflectivity was measured using an amplified Si photodiode and lock-in detection.

\section*{Results and Discussion}

\subsection*{Growth and Structural Characterization}

In studying the epitaxial growth of FGT on Ge(111), we observe a fascinating dependence of the materials quality on the growth rate. To illustrate, we employ two representative growth rates, namely a ``Slow Growth'' rate of $\sim 0.06$ \AA/s and a ``Fast Growth'' rate of $\sim 0.12$ \AA/s.
Starting from streaky RHEED patterns of the Ge(111) substrate (Figure~\ref{fig:Ge_RHEED}), in both cases, we observe streaky RHEED throughout the growth as shown in Figures~\ref{fig:slow_FGT_RHEED} and \ref{fig:fast_FGT_RHEED}.
In addition, the streak spacing matches the FGT in-plane lattice constant in both cases. However, the X-ray diffraction (XRD) shows different features for Slow Growth and Fast Growth, even though their RHEED patterns are very similar. For Slow Growth FGT in Figure~\ref{fig:slow_FGT_XRD}, it can be seen that the $\theta - 2\theta$ XRD scan has an unexpected extra peak of unknown origin (red arrow) beside the FGT(006) peak. But for Fast Growth FGT in Figure~\ref{fig:fast_FGT_XRD}, the extra peak in the XRD scan no longer exists. This indicates a pure phase FGT film for Fast Growth and a mixed phase FGT film for Slow Growth.

\begin{figure}[t]
  \adjustbox{minipage=1.1em,valign=b}
  \centering
    \includegraphics[width=0.95\textwidth,valign=t]{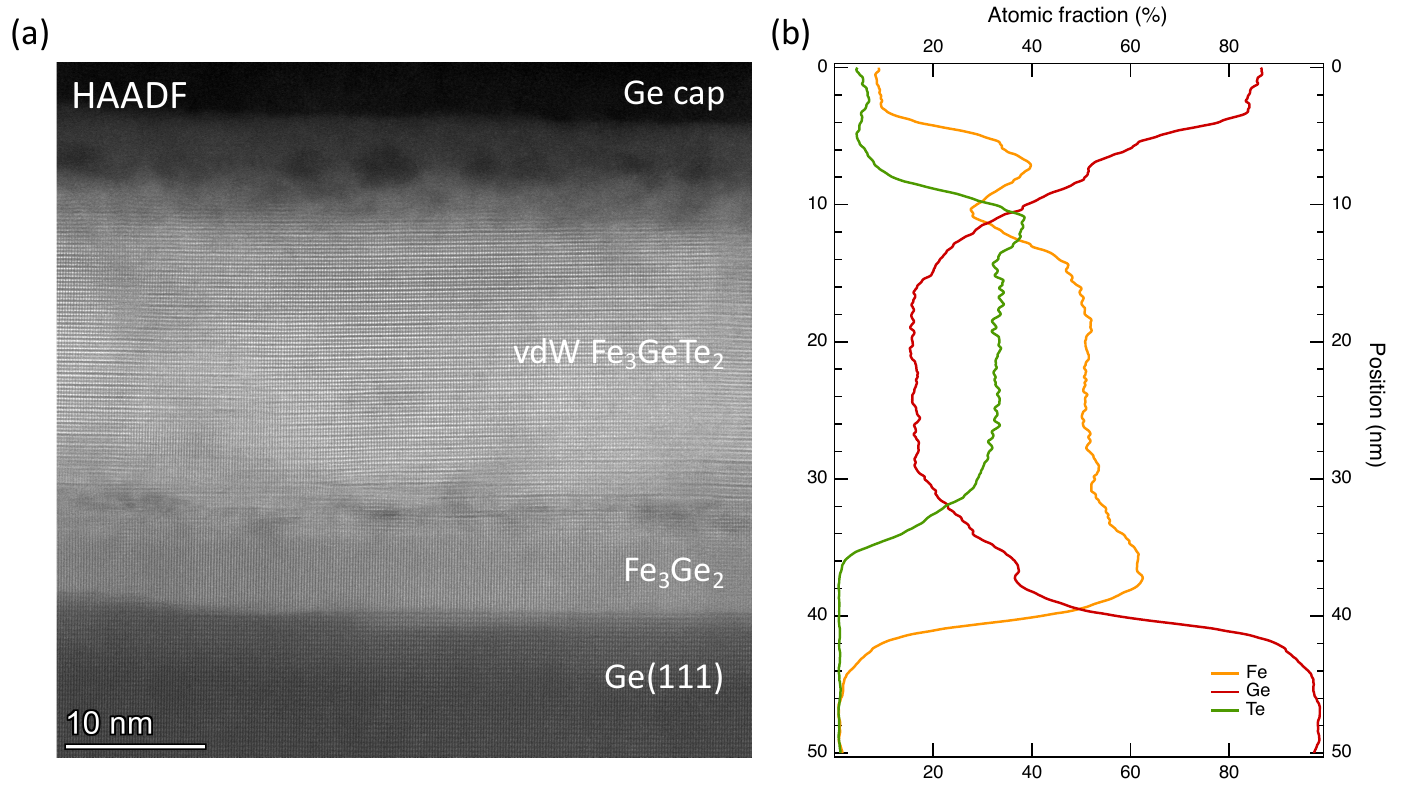}
\hfill
\caption{Scanning transmission electron microscopy (STEM) of Slow Growth FGT on Ge(111). (a) Cross-sectional STEM image of Slow Growth FGT ($\sim 0.06$ \AA/s). (b) Energy dispersive x-ray spectroscopy (EDX) of Slow Growth FGT as a function
of depth, showing Te deficiency during the initial growth. \label{fig:slow_TEM_XEDS}
}
\end{figure}

To better understand this phenomenon, we perform cross-sectional STEM for atomic-scale characterization of the structure and chemical composition. For the Slow Growth FGT (Figure~\ref{fig:slow_TEM_XEDS}a), we observe a non-vdW phase at the initial growth on the Ge substrate. But good vdW layers subsequently grow on top of this initial non-vdW phase. The vdW layers are clearly visible in the STEM image by the bright horizontal lines. 
The chemical composition shown in the EDX scan (Figure~\ref{fig:slow_TEM_XEDS}b) provides some key insights. In the top part of the film, the stoichiometry of Fe, Ge and Te is 3.13:1.00:2.03 (averaged in between position 20 nm to 26 nm) indicating the formation
of the vdW magnet Fe$_3$GeTe$_2$. In the bottom
part of the film, the Te atomic fraction goes to zero and the Fe and Ge atoms have a ratio of $\sim$1.56 (averaged in between position 36 nm to 39 nm), which suggests the alloy Fe$_3$Ge$_2$.
This conclusion is further supported by the extra peak in the XRD corresponding to a lattice spacing of $\sim$5.04 \AA, which
is close to the lattice spacing of 5.010 \AA~\cite{vaughn_solution-phase_2013,goswami_growth_2005} for Fe$_3$Ge$_2$ along the [001] direction (hexagonal c-plane). In addition, Fe$_3$Ge$_2$ has an in-plane lattice constant of $a=b=3.998$ \AA~\cite{vaughn_solution-phase_2013,goswami_growth_2005}, which is very similar to that of FGT ($a=b=3.991$ \AA~\cite{roemer_robust_2020,deiseroth_fe3gete2_2006,yi_competing_2016}). This explains why it is hard to tell the difference between FGT and Fe$_3$Ge$_2$ in RHEED during growth. Thus, the EDX, XRD, and RHEED provide strong evidence that the non-vdW interfacial phase is (001)-oriented Fe$_3$Ge$_2$.

\begin{figure}
  \adjustbox{minipage=1.1em,valign=t}{\subcaption{}\label{fig:fast_FGT_TEM}}
  \begin{subfigure}[t]{\dimexpr.85\textwidth-1.1em\relax}
  \centering
    \includegraphics[width=0.95\textwidth,valign=t]{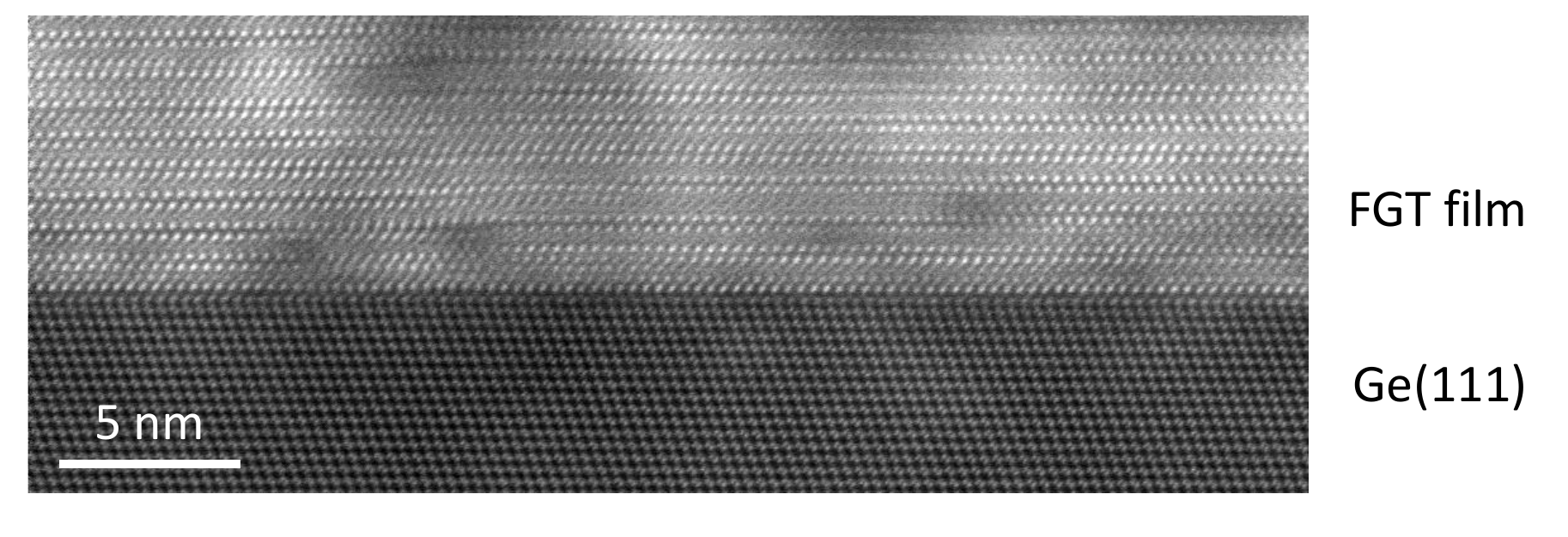}
    \end{subfigure}%
\hfill
  \adjustbox{minipage=1.1em,valign=t}{\subcaption{}\label{fig:fast_FGT_STM_step}}%
  \begin{subfigure}[t]{\dimexpr.45\textwidth-1.1em\relax}
  \centering
    \includegraphics[width=0.95\textwidth,valign=t]{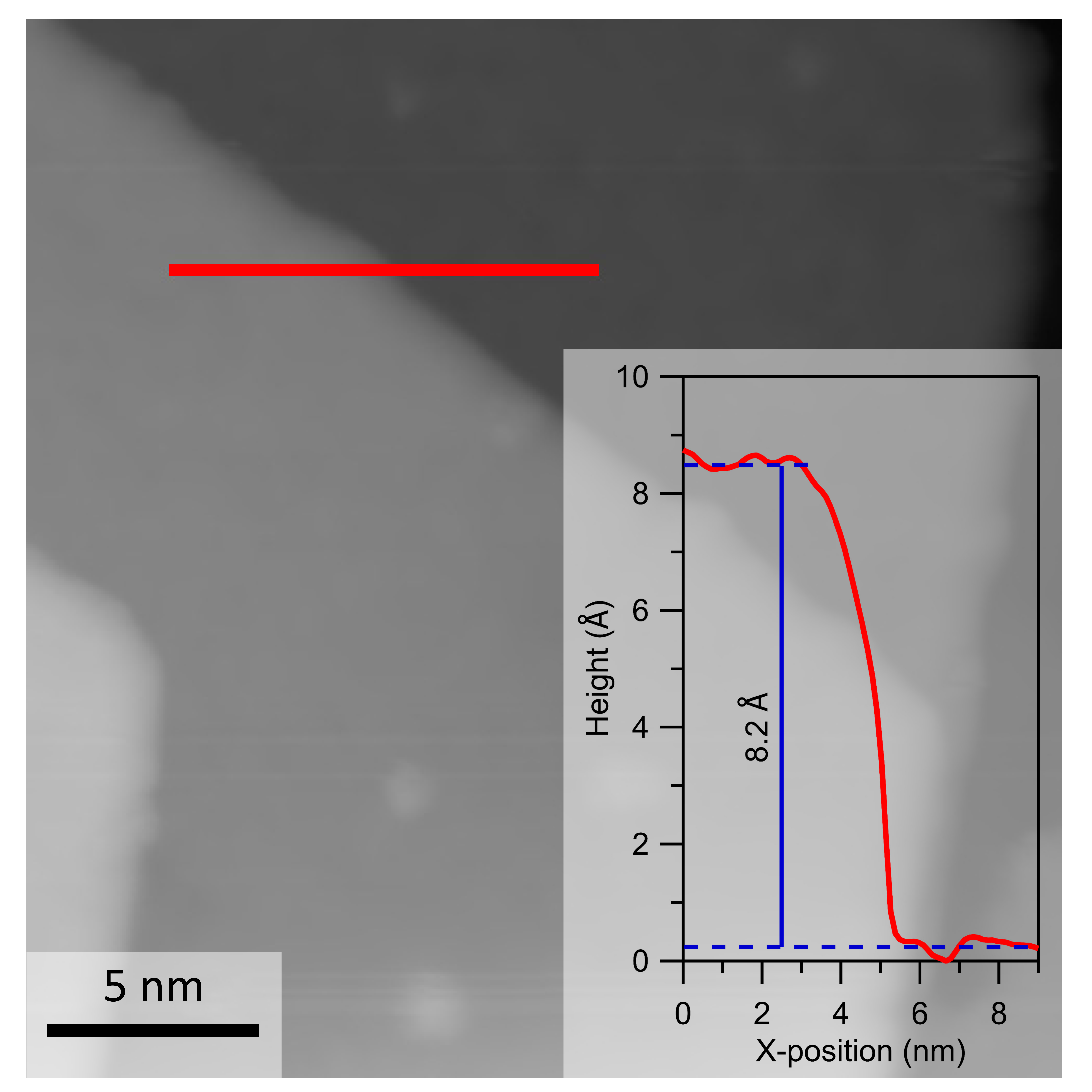}
    \end{subfigure}
  \hfill
  \adjustbox{minipage=1.1em,valign=t}{\subcaption{}\label{fig:fast_FGT_STM_atomic}}%
  \begin{subfigure}[t]{\dimexpr.45\textwidth-1.1em\relax}
  \centering
    \includegraphics[width=0.95\textwidth,valign=t]{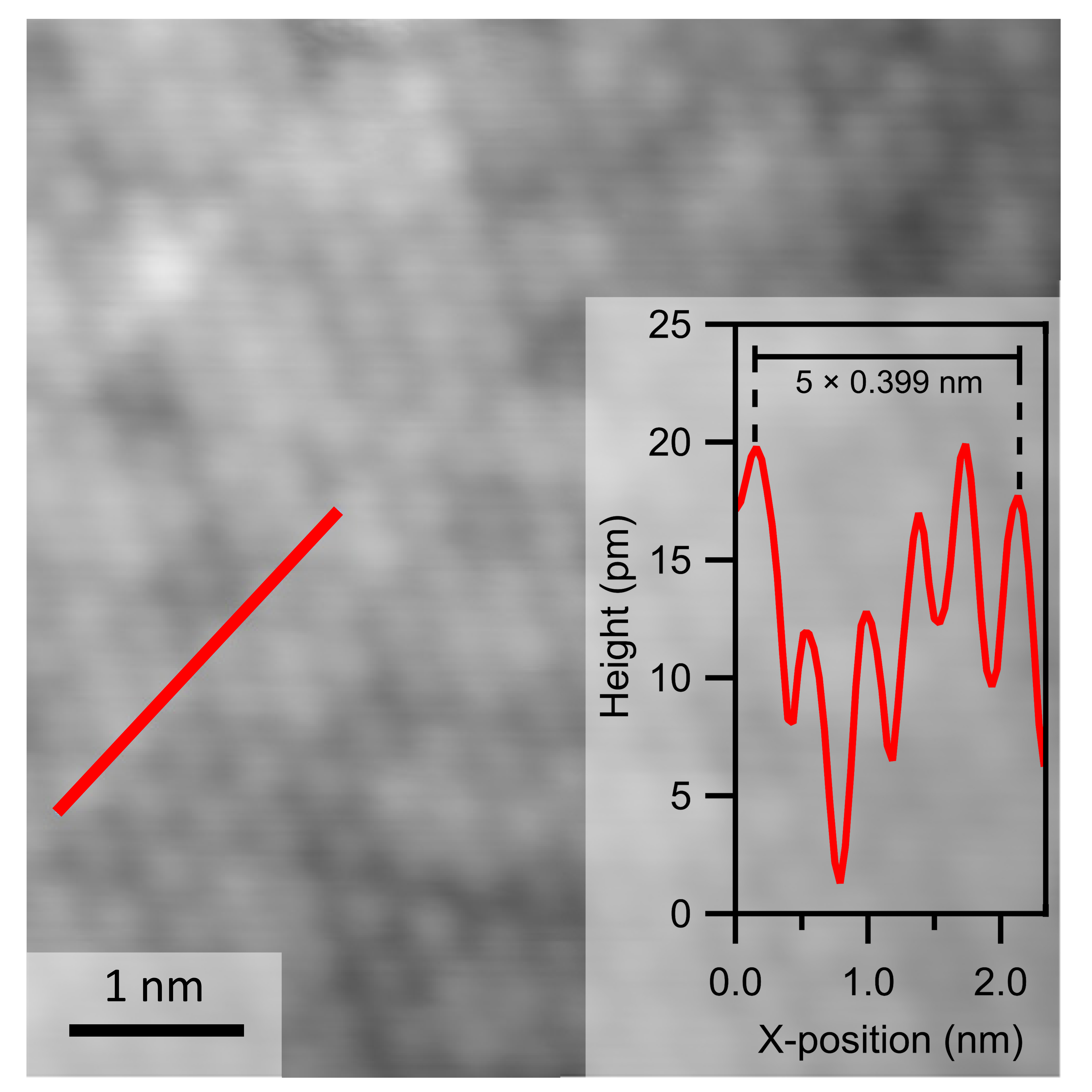}
    \end{subfigure}
  \hfill
\caption{Atomic scale structural characterization of Fast Growth FGT on Ge(111). (a) Cross-sectional STEM image of Fast Growth FGT showing atomically abrupt interfaces between the van der Waals FGT and the Ge substrate. (b) Surface morphology of Fast Growth FGT measured by STM. Inset: Height profile line cut across a step-edge indicated by the solid red line (tunneling conditions: sample bias V = -154 mV, I = 0.22 nA). (c) Atomically-resolved imaging of Te terminated FGT measured by STM. Inset: line cut profile indicated by the solid red line with a periodicity of 3.99 \AA~(tunneling conditions: sample bias V = 100 mV, I = 0.15 nA).  \label{fig:fast_TEM_STM}
}
\end{figure}

The Fast Growth FGT on Ge(111) exhibits quite different behavior. As shown in Figure ~\ref{fig:fast_FGT_TEM}, cross-sectional STEM imaging reveals that vdW FGT layers form immediately at the start of growth. Again, the vdW layers are clearly visible in the TEM by the bright horizontal lines.
Moreover, a close look at the interface shows an atomically sharp transition between the Ge substrate and the FGT film, which indicates that the vdW FGT layers form immediately at the interface. This is a great improvement over the Slow Growth FGT (Figure~\ref{fig:slow_TEM_XEDS}) which has a transition layer of Fe$_3$Ge$_2$ in between the Ge substrate and the FGT film. However, there is still room for improvement, as some defects are visible.

STM measurements on a Fast Growth FGT sample, shown in Figures~\ref{fig:fast_FGT_STM_step} and \ref{fig:fast_FGT_STM_atomic}, also support the high quality of films grown by this method.
Figure~\ref{fig:fast_FGT_STM_step} shows large atomically flat terraces with a line cut across a step edge.
The height profile of the line cut, displayed in the inset, indicates that the two terraces differ in height by 8.2 \AA , which is in good agreement with the height of a single vdW layer of FGT.
Atomic resolution imaging (Figure~\ref{fig:fast_FGT_STM_atomic}) reveals a well ordered hexagonal lattice with an in-plane lattice constant of 3.99 \AA , consistent with the topmost Te layer of an FGT vdW layer.
These results indicate that the growth method established here facilitates good layer-by-layer growth with consistent Te termination.

\begin{figure}
  \adjustbox{minipage=1.1em,valign=t}{\subcaption{}\label{fig:fast_FGT_AHE}}
  \begin{subfigure}[t]{\dimexpr.4\textwidth-1.1em\relax}
  \centering
    \includegraphics[width=\textwidth,valign=t]{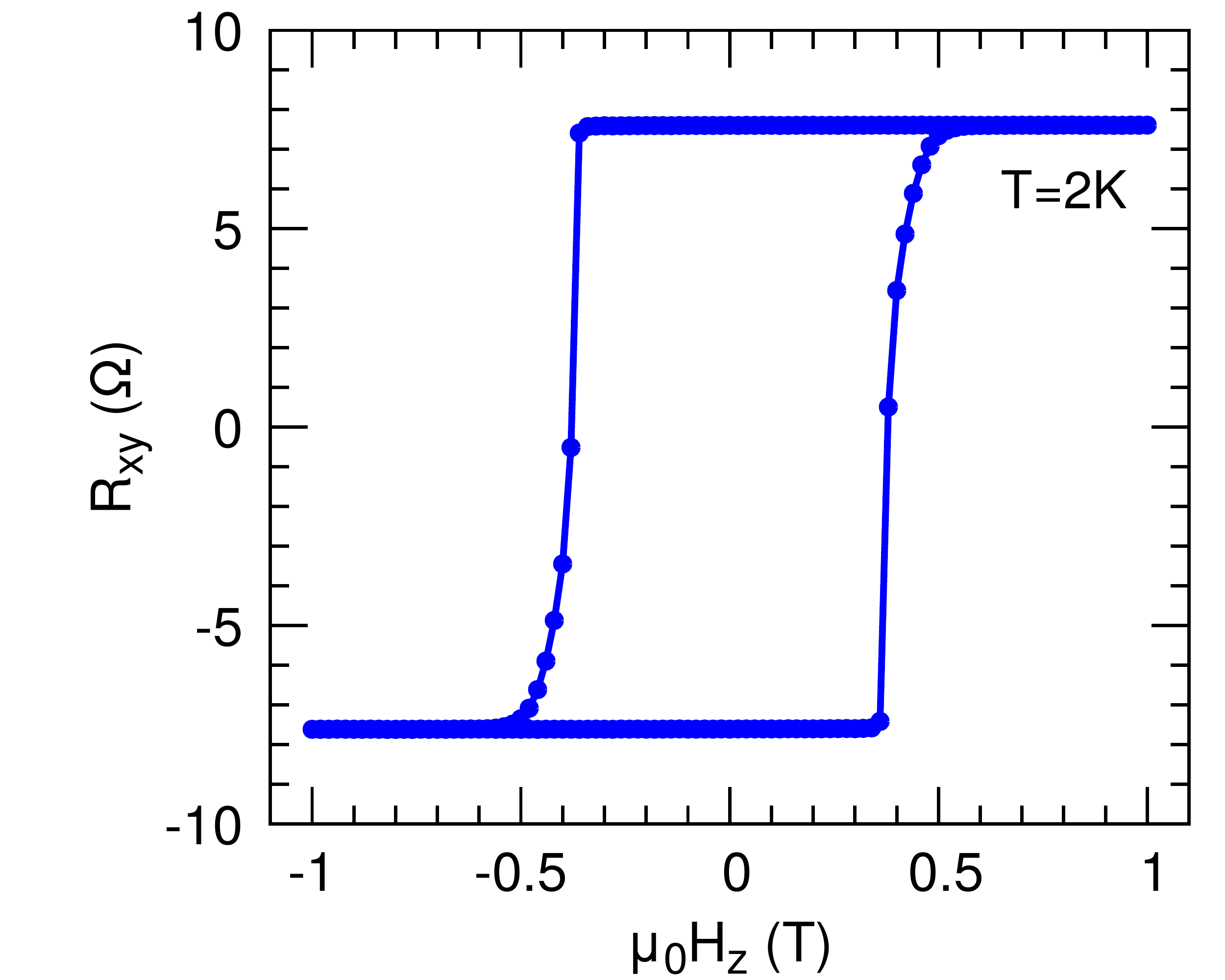}
    \end{subfigure}%
  \adjustbox{minipage=1.1em,valign=t}{\subcaption{}\label{fig:fast_FGT_MCD}}
  \begin{subfigure}[t]{\dimexpr.4\textwidth-1.1em\relax}
  \centering
    \includegraphics[width=\textwidth,valign=t]{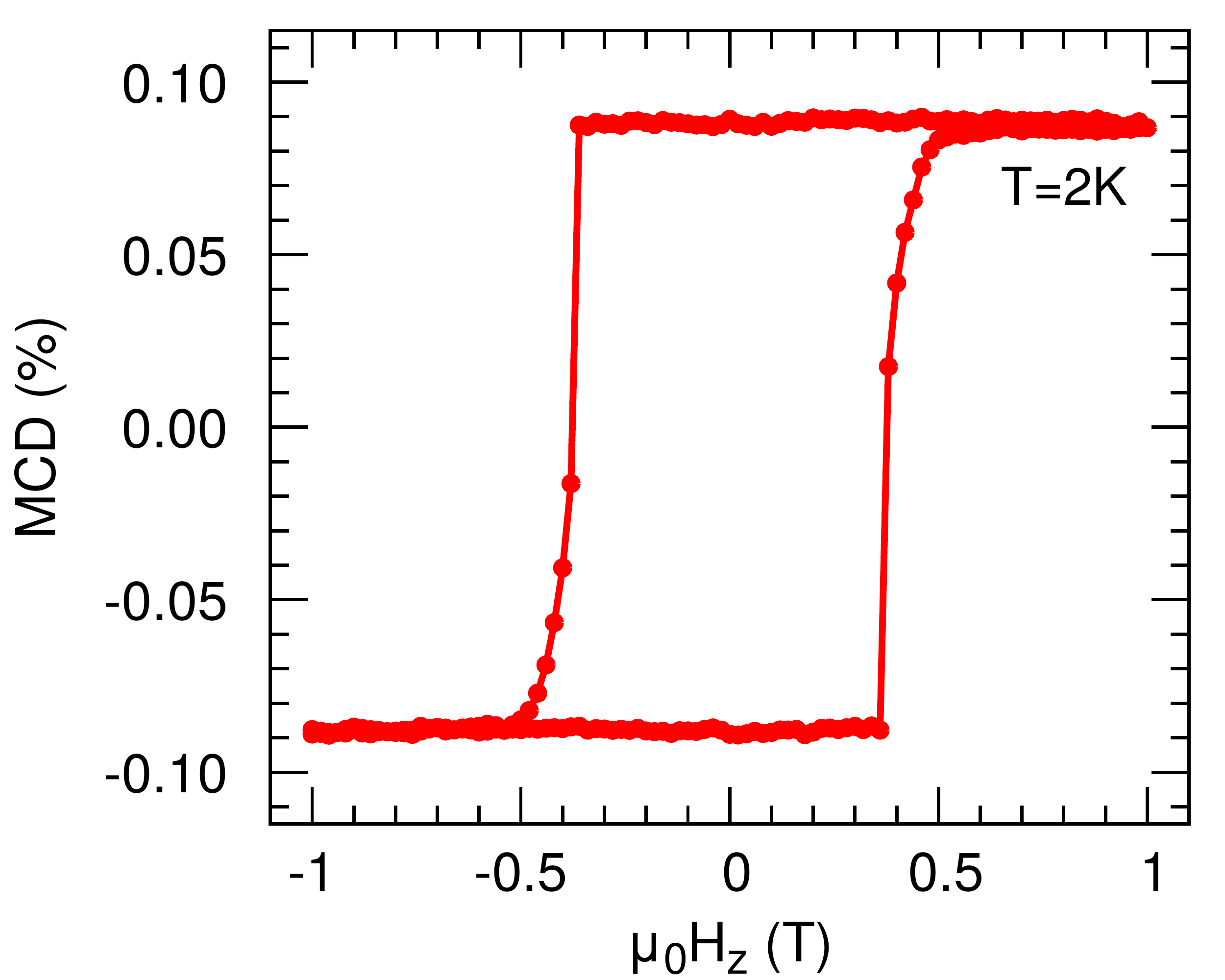}
    \end{subfigure}%
\hfill
  \adjustbox{minipage=1.1em,valign=t}{\subcaption{}\label{fig:fast_FGT_MCD_T}}%
  \begin{subfigure}[t]{\dimexpr.5\textwidth-1.1em\relax}
  \centering
    \includegraphics[width=\textwidth,valign=t]{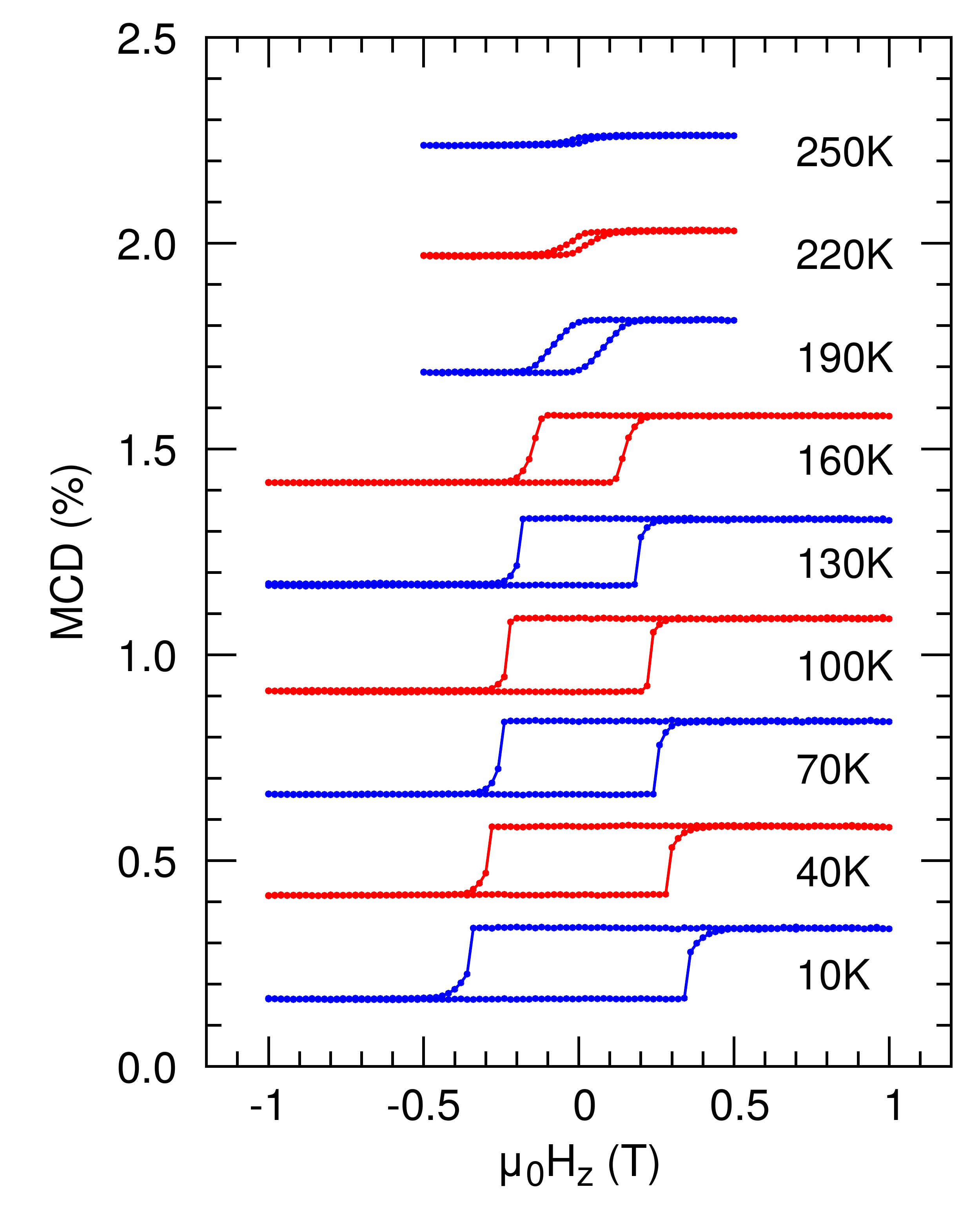}
    \end{subfigure}
  \hfill
  \adjustbox{minipage=1.1em,valign=t}{\subcaption{}\label{fig:fast_FGT_SPSTM}}%
  \begin{subfigure}[t]{\dimexpr.55\textwidth-1.1em\relax}
  \centering
    \includegraphics[width=\textwidth,valign=t]{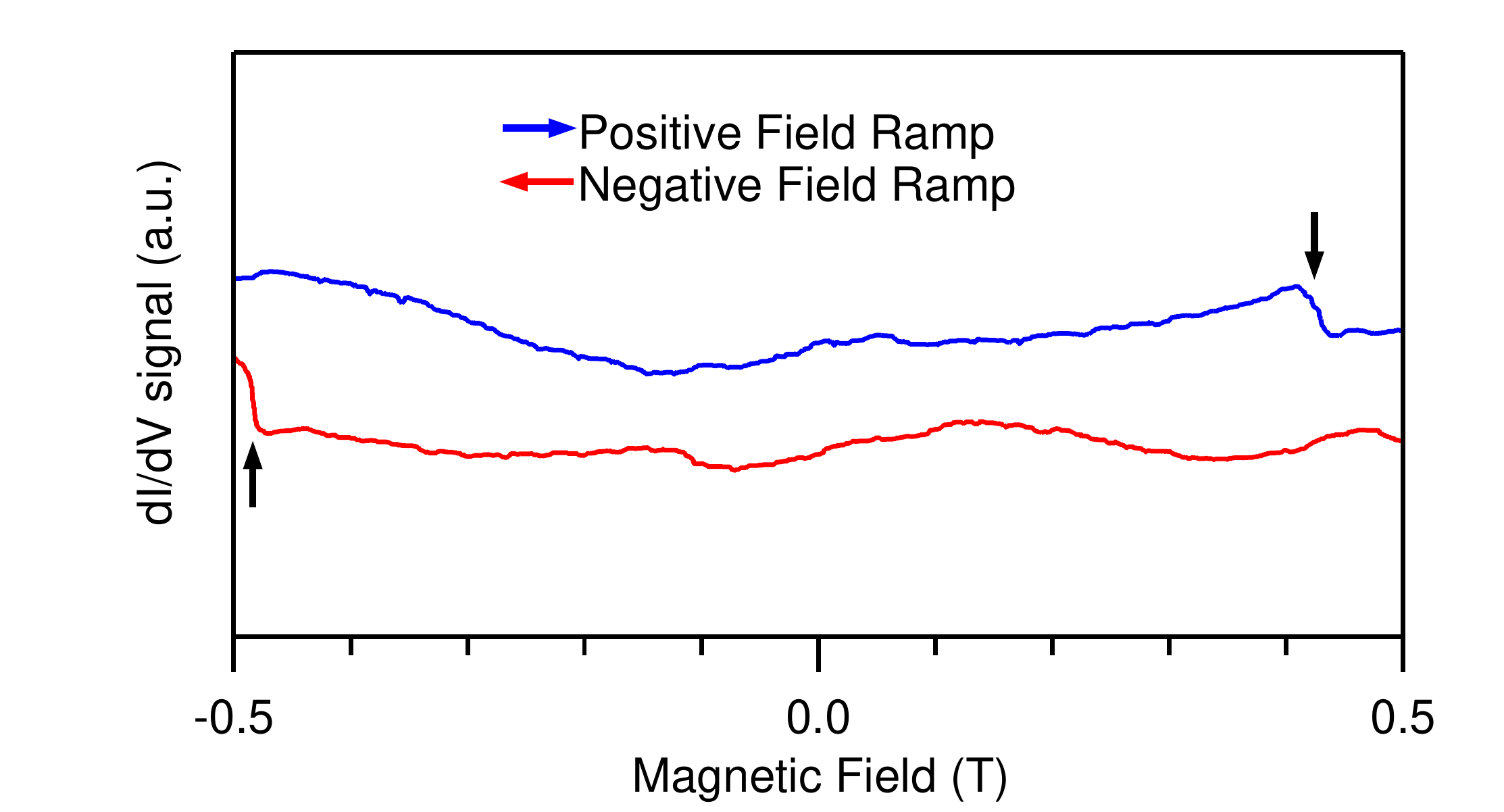}
    \end{subfigure}
  \hfill
\caption{ Out-of-plane magnetic hysteresis loops of Fast Growth FGT. (a) Anomalous Hall effect measurement at 2 K. (b) Magnetic circular dichroism (MCD) measurement at 2 K. (c) MCD for a series of temperatures, showing a T$_C$ of $\sim$250 K. Curves have been vertically offset. (d) Spin-polarized scanning tunneling microscopy measurement at 5 K. The data has been filtered for noise reduction. Curves have been offset for clarity.\label{fig:magnetic hysteresis}}
\end{figure}

\subsection*{Magnetic Properties}

For the Fast Growth FGT/Ge(111) samples, we investigate the out-of-plane hysteresis loops by measuring the anomalous Hall effect (AHE), reflection magnetic circular dichroism (MCD), and spin-polarized STM.  Figures~\ref{fig:fast_FGT_AHE} and~\ref{fig:fast_FGT_MCD} show out-of-plane hysteresis loops of a 20 nm Fast Growth FGT film measured by AHE and MCD, respectively, at 2 K. 
The AHE measurement exhibits a square hysteresis loop and a coercivity of $\sim$0.379 T, similar to values reported in the literature. 
Similarly, the hysteresis loop measured by MCD exhibits virtually identical characteristics as the AHE hysteresis loop measured in the same cryostat. 
The temperature dependence of the MCD hysteresis loops is shown in Figure~\ref{fig:fast_FGT_MCD_T}. From 10 K to 130 K, the hysteresis loops remain square and the coercivity reduces from 0.355 T to 0.199 T. 
From 160 K to 250 K, the magnetization reversal becomes more gradual and the coercivity continues to decrease. 
The magnitude of the MCD signal reduces to nearly zero at 250 K, leading to an estimated T$_C$ of $\sim$250 K. These magnetic properties reflect the high quality of the epitaxial films.

Spin-polarized STM measurements\cite{repicky2021} are performed using a Cr tip and sweeping the out-of-plane magnetic field. The magnetic constrast is observed in the dI/dV signal at a sample bias of 100 mV and a tip current of 0.15 nA.
As shown in Figure~\ref{fig:fast_FGT_SPSTM}, the dI/dV signal shows sharp switchings indicated by black arrows at 0.42 for the positive field ramp (blue) and at -0.48 T for the negative field ramp (red). 
Unlike the MCD measurement which averages the magnetization within the $\sim$150\,$\mu$m laser spot, the STM measurement is local at the atomic scale. 
This explains why MCD hystersis loops are more uniform while the STM hysteresis loop has switchings that occur at different field magnitudes. 
The latter reflects the stochastic nature of magnetization reversal. Future studies are planned to capture images of the magnetic domain structure during a magnetization reversal.

\subsection*{Discussion}

We now consider why growth rate is such an important factor in determining the material quality during the initial growth.
The fact that the synthesis depends strongly on growth rate implies the importance of kinetics, or the rate of material processes (e.g. interdiffusion, chemical reaction, atomic diffusion on surfaces, re-evaporation, etc.). 
The complementary factor that is significant for material synthesis is the thermodynamic equilibrium phase which minimizes the free energy.
The balance between energetics and kinetics can be controlled by the growth rate, where slower growth rates favor the equilibrium phase and faster growth rates can produce materials further away from equilibrium based on kinetic considerations.

In applying these concepts to our case, we first point out that the growth conditions with a 10$\times$ overpressure of Te (i.e.~Fe:Ge:Te beam flux ratio of 3:1:20 for Fe$_3$GeTe$_2$ with a stoichiometry of 3:1:2) is a typical condition for reactive MBE, where the volatile species (Te) reacts with the other elements to form the desired material with proper stoichiometry due to re-evaporation of excess Te.
In reactive MBE, when the film has a deficiency of the volatile species, typical solutions are to increase its beam flux or reduce the growth rate. This allows more time for the reaction to occur and can increase the incorporation of the volatile species into the material. 
But our FGT growth appears to behave in an opposite way, so we would not consider this process as traditional reactive MBE. 
Since slower growth favors the equilibrium phase while faster growth is further from equilibrium, this suggests that a Fe$_3$Ge$_2$ surface is more thermodynamically stable than the FGT surface within a partial pressure of $10^{-8}$ to $10^{-7}$ Torr of Te and at 325 $\degree$C during the intial stages of growth.
In other words, energetics favor formation of Fe$_3$Ge$_2$. Thus, in order to stabilize the FGT phase that we want, we must consider the rate kinetics associated with the growth. 
In comparing the chemical composition of Fe$_3$Ge$_2$ with Fe$_3$GeTe$_2$, the Fe$_3$Ge$_2$ has a higher relative content of Ge (Ge:Fe = 2:3 for Fe$_3$Ge$_2$ vs.~1:3 for FGT) and an absence of Te.
The absence of Te in Fe$_3$Ge$_2$ is achieved by the out-diffusion of Te from the film into vacuum. For the increased Ge:Fe ratio (2:3) relative the beam flux ratio (1:3), either the sticking coefficient of the Fe must become lower than that of Ge or the excess Ge is supplied by the substrate. 
Since the cell temperature for Fe of $\sim$1300 $\degree$C is much higher than the substrate temperature of 325 $\degree$C, this argues against a reduced sticking coefficient of Fe and strongly supports the scenario that the excess Ge is supplied via interdiffusion from the substrate. 
This is also supported by the fact that the growth eventually converts to a vdW FGT phase.
Thus, to stabilize the FGT phase we want (and avoid Fe$_3$Ge$_2$), the growth rate must be sufficiently high to outpace the out-diffusion of Te from the film surface and the interdiffusion of Ge into the film. 
One would expect the required minimum growth rate to be temperature dependent. Lower temperatures have slower diffusion so it could tolerate slower growth rates, while higher temperatures would require faster growth rates. 
Considering that the optimized growth temperatures are usually a balance between maximizing smoothness and reducing defects (favoring higher T) and minimizing interdiffusion (favoring lower T), our current results indicate that growth rate should also be strongly considered in the materials optimization.

Identifying this kinetically-controlled growth regime for FGT is important for the future optimization of the material and its incorporation into epitaxial heterostructures. While the Fe$_3$Ge$_2$ phase might be specific to FGT on a Ge substrate, the issues of interdiffusion and chemical reaction with a substrate or underlayer are universally important for film growth. 
Our results suggest that the FGT growth rate may be used to control interdiffusion and chemical reaction with underlayers more generally, which will be useful for incorporating FGT into epitaxially-grown vdW heterostructures.

\section*{Conclusion}

In summary, we discovered that an important factor for optimizing FGT synthesis is the growth rate, which has an unexpected influence on thin film quality.
The phenomenon is explained by energetics and kinetics of MBE growth where the stabilization of FGT is assisted by a higher deposition rate that inhibits the non-vdW Fe$_3$Ge$_2$ phase from forming.
Atomic-scale imaging by STM and STEM confirm the structural quality of the surface and interface.
The excellent magnetic properties with square out-of-plane hysteresis loops is confirmed by AHE, MCD, and spin-polarized STM measurements.
Our results reveal an important way to think about and optimize MBE growth, leading to potentially better 2D materials.


\begin{acknowledgement}

We thank Shuyu Cheng, Katherine Robinson, and Ryan Bailey-Crandell for technical assistance. This work was supported by AFOSR MURI 2D MAGIC Grant No.~FA9550-19-1-0390 (MBE, magneto-optics), U.S. DOE Office of Science, Basic Energy Sciences Grant No.~DE-SC0016379 (STM), and the Center for Emergent Materials, an NSF MRSEC, under Grant No.~DMR-2011876 (TEM, magnetotransport). Electron microscopy was performed at the Center for Electron Microscopy and Analysis (CEMAS) at The Ohio State University.

\end{acknowledgement}




\section*{Author Contributions Statement}

W.Z., A.J.B., and R.K.K. conceived the experiments.  
W.Z. and A.J.B. performed the MBE growth. M.Z. and J.H. performed the TEM measurements. A.J.B., R.C.W., and J.A.G. performed the STM measurements. I.L., W.Z., and R.K.K. performed the magnetotransport and magnetooptic measurements.
All authors participated in data analysis and preparation of the manuscript.


\bibliography{main.bib}



\end{document}